\documentclass[floatfix,english,aps,prb,twocolumn,amsmath,amssymb]{revtex4}

\usepackage{color}
\usepackage{graphicx}
\usepackage[caption=false]{subfig}
\usepackage[version=4]{mhchem}
\graphicspath{{.}{./figures/}}

\usepackage{siunitx}

\usepackage{multirow}
\usepackage[utf8x]{inputenc}

\usepackage{upgreek}
\AtBeginDocument{}
\usepackage[capitalise,noabbrev]{cleveref}
\Crefname{equation}{reaction}{reactions}
\usepackage{xr}
\newcommand*{\etal}{\textit{et~al.}}

\newcommand*{\Hreac}{$\Delta_{r}H^\ominus$}
\newcommand*{\Greac}{$\Delta_{r}G^\ominus$}
\newcommand*{\Hvap}{$\Delta_{vap}H^\ominus$}
\newcommand*{\Hform}{$\Delta_fH^\ominus$}
\newcommand*{\Tref}{\SI{298.15}{\kelvin}}

\begin{document}

\title{Facing the challenge of predicting the standard formation enthalpies of n-butyl-phosphate species with ab initio methods}

\author{Mohamad Saab}
\affiliation{Laboratoire PhLAM, CNRS UMR 8523, Université de Lille, 59655 Villeneuve d'Ascq Cedex, France}
\author{Florent Réal}
\affiliation{Laboratoire PhLAM, CNRS UMR 8523, Université de Lille, 59655 Villeneuve d'Ascq Cedex, France}
\author{Martin \v{S}ulka}
\affiliation{Slovak University of Technology in Bratislava, ATRI, Faculty of Materials Science and Technology in Trnava, Bottova 25, 917 24 Trnava, Slovak Republic}
\author{Laurent Cantrel}
\affiliation{Institut de Radioprotection et de Sûreté Nucléaire (IRSN), PSN-RES, Cadarache, St Paul Lez Durance, 13115, France}
\author{François Virot}
\affiliation{Institut de Radioprotection et de Sûreté Nucléaire (IRSN), PSN-RES, Cadarache, St Paul Lez Durance, 13115, France}
\author{Valérie Vallet}
\email{valerie.vallet@univ-lille1.fr}
\affiliation{Laboratoire PhLAM, CNRS UMR 8523, Université de Lille, 59655 Villeneuve d'Ascq Cedex, France}
\date{\today}

\begin{abstract}
Tributyl-phosphate (\ce{TBP}), a ligand used in the PUREX liquid-liquid separation process of spent nuclear fuel, can form explosive mixture in contact with nitric acid, that might lead to violent explosive thermal runaway. In the context of safety of a nuclear reprocessing plant facility, it is crucial to predict the stability of TBP at elevated temperatures. So far, only the enthalpies of formation of TBP is available in the literature with a rather large uncertainties, while those of its degradation products, di-(\ce{HDBP}) and mono-(\ce{H2MBP}) are unknown. In this goal, we have used state-of-the art quantum chemical methods to compute the formation enthalpies and entropies of TBP and its degradation products di-(\ce{HDBP}), mono-(\ce{H2MBP}) in gas and liquid phases. Comparisons of levels of quantum chemical theory revealed that there are significant effects of correlation on their electronic structures, pushing for the need of not only high level of electronic correlation treatment, namely local coupled cluster with single and double excitation operators and perturbative treatment of triple excitations [LCCSD(T)], but also extrapolations to the complete basis to produce reliable and accurate thermodynamics data. Solvation enthalpies were computed with the conductor like screening model for real solvents [COSMO-RS], for which we observe errors not exceeding \SI{22}{\kJ\per\mol}. We thus propose with final uncertainty of about \SI{20}{\kJ\per\mol} standard enthalpies of formation of \ce{TBP}, \ce{HDBP}, and \ce{H2MBP} which amounts to \num{-1281.7\pm24.4}, \num{-1229.4\pm19.6} and \SI{-1176.7\pm14.8}{\kJ\per\mol}, respectively, in the gas phase. In the liquid phase, the predicted values are \num{-1367.3\pm24.4}, \num{-1348.7\pm19.6} and  \SI{-1323.8\pm14.8}{\kJ\per\mol}, to which we may add about \SI{-22}{\kJ\per\mol} error from the COSMO-RS solvent model. From these data, the complete hydrolysis of \ce{TBP} is predicted as an exothermic phenomena but showing a slightly endergonic process. 
\end{abstract}
\maketitle

\section{Introduction}

All spent nuclear fuel reprocessing plants use the PUREX process  (Plutonium Uranium Refining by Extraction), which is a liquid-liquid extraction method. The organic extracting solvent is a mixture of tri-n-butyl phosphate (TBP) and hydrocarbon solvent such as hydrogenated tetra-propylene (TPH). By chemical complexation, uranium and plutonium (from spent fuel dissolved in nitric acid solution), are separated from fission products and minor actinides.  During a normal extraction operation, uranium is extracted in the organic phase as the \ce{UO2(NO3)2(TBP)2} complex.

The TBP extractant can form an explosive mixture called red oil when it comes in contact with nitric acid. The formation of this unstable organic phase originates from the reaction between TBP and its degradation products on the one hand, and nitric acid, its derivatives and heavy metal nitrate complexes on the other hand. The decomposition of the red oil can lead to violent explosive thermal runaway. These hazards are at the origin of several accidents such as, the two in the United States in 1953 and 1975 (Savannah River) \cite{Colven53,Durant83}  and, more recently, the one in Russia in 1993 (Tomsk). \cite{IAEA98} The accidental consequences were the rupture of the equipment, followed by a significant radioactive release to the environment. This raises the question of the exothermicity of reactions that involve TBP and all other degradation products, and calls for a better knowledge of the underlying chemical phenomena. 

TBP is relatively stable, but at elevated temperatures, its degradation may occur, as a result of the following phenomena: radiolysis, pyrolysis,  acid catalyzed hydrolysis, as well as the de-alkylation (see the therein reference for a detailed reaction scheme.\cite{Wright20101753}) The two last lead to the stepwise formation of degraded organic products such as dibutyl phosphate (HDBP), mono-butyl phosphate (\ce{H2MBP}), and ultimately phosphoric acid (\ce{H3PO4}). The reaction by-products (butanol and butyl nitrate) may then undergo oxidation to form carboxylic acids and gases. In a similar way, \ce{UO2(NO3)2(TBP)2} may be decomposed to \ce{UO2(NO3)2TBP(HDBP)} or \ce{UO2(NO3)2(HDBP)2}. To understand the chemistry of these systems, the knowledge of some physical properties of such species is necessary, such as thermodynamic functions or kinetics rate law of degradation. However, such data are rather scarce in the available literature. Even if there are a plethora of studies on TBP-nitric acid systems (see reviews \onlinecite{Wright20101753,thermo-Burger-Book1990}), some of the fundamental properties, like standard enthalpies of formation or kinetics of by-product degradations, are poorly known. 

Only three articles have reported experimental investigations on standard enthalpy of formation of TBP  and there is no data for degraded organic products. These published data were obtained by combustion method in a bomb calorimeter. Starostin~{\etal}\cite{actinide-Starostin-IANSSK1966-15-1255} first studied the TBP combustion with static isothermal calorimeter, the value, \SI{-1458\pm12}{\kJ\per\mol} was obtained. Later, Kindle\cite{actinide-Kindle-techreport1974} reported measurements carried on adiabatic bomb calorimeter targeting the amount of heat of combustion in a nuclear facility for safety analysis. Even if initially the data are not intended for the calculation of enthalpy of formation, an evaluation is possible by supposing a complete combustion:  \SI{-1250\pm69}{\kJ\per\mol}. The most recent study, carried out by Erastov and Tarasov\cite{physchem-Erastov-RJPC1992-66-2591} derived from rotating bomb calorimeter measurements, the standard enthalpy of formation of TBP; the value \SI{-1382\pm12}{\kJ\per\mol} was retained.  Despite similar approaches, experimental data exhibit a large uncertainty over a range of about \SI{200}{\kJ\per\mol}. Therefore the aim of our safety strategy of red oil issues is split into several steps where the first investigation is the calculation of standard enthalpy of formation of TBP in the TBP solvent as well as its degraded products, HDBP and \ce{H2MBP}. 

In the first part of this article, the quantum chemistry calculations are introduced. The choice of electronic correlation treatment and a benchmark of basis sets are discussed as the methodology to evaluate the solvent contributions. In the second part, the standard enthalpy of formation of target species in both gas and liquid phases, as well as the hydrolysis reaction free energies, are presented and discussed regarding the experimental data.

\section{Computational Details}\label{sec:comp}

\subsection{Chemical reactions used to derive the standard enthalpies of formation of \ce{H_{3-n}nBP}}
The standard enthalpies of formation in the gas and TBP liquid phases of the \ce{H_{3-n}nBP} molecules, \ce{H2MBP} (n=1), \ce{HDBP} (n=2), and \ce{TBP} (n=3) are derived from computed enthalpies of reactions. 

\begin{widetext}
\begin{align}
\ce{H3PO4(l) + n C4H9OH(g) & -> H_{3-n}nBP(g)/(l) +  n H2O(g),\label{rxn:H3PO4}\\
POF3(g) + n C4H9OH(g) + (3 - n) H2O(g) & -> H_{3-n}nBP(g)/(l) + 3 HF(g)\label{rxn:HF}\\
POCl3(g) + n C4H9OH(g) + (3 - n) H2O(g) &-> H_{3-n}nBP(g)/(l) + 3 HCl(g), \label{rxn:HCl}\\
PH3(g) + n C4H9OH(g) + (4 - n) H_2O(g) &-> H_{3-n}nBP(g)/(l)+ 4 H2(g).\label{rxn:H2}}
\end{align}
\end{widetext}

together with the standard enthalpies of formation reported in~\cref{tab:Hfdata}.
\begin{table}
\caption{\label{tab:Hfdata}Known Gas/Liquid Phase Standard Enthalpies of Formation {\Hform}({\Tref}) in \si{\kJ\per\mol}}
\begin{ruledtabular}
\begin{tabular}{*2lc}
{Molecule} & {{\Hform}({\Tref})} & {Reference}\\
\hline
\ce{HF(g)}		&\num{-273.3(7)} 		& {\citenum{nist-Cox-Book1984}}\\
\ce{HCl(g)}		&\num{-92.31(10)} 		& {\citenum{nist-Cox-Book1984}}\\
\ce{H2O(g)}		& \num{-241.826(040)} 	& {\citenum{nist-Cox-Book1984}}\\
\ce{C4H9OH(g)}& \num{-277(5)} 		& {\citenum{NIST-Webbook}}\\
\ce{PH3(g)} 	& \num{5.47\pm1.7} 			& {\citenum{nist-Chase-JPCRD1998--1-1951}}\\
\ce{POF3(g)}& \num{-1254.25\pm 8} 		& {\citenum{nist-Chase-JPCRD1998--1-1951}}\\
\ce{POCl3(g)}	& \num{-559.82\pm1.7} 		& {\citenum{nist-Chase-JPCRD1998--1-1951}}\\
\ce{H3PO4(l)}		& \num{-1271.66\pm2.9}\footnote{Details on how the reported uncertainty was estimated are given in the supplementary material. Further comparisons with computed gas-phase formation enthalpies are also included in supplementary material}	& {\citenum{nist-Cox-Book1984}}\\
\end{tabular}
\end{ruledtabular}

\end{table}

Note that all these reactions involve closed-shell molecules. As discussed by Watts\cite{thermo-Watts-Book2016} and illustrated by Hehre\cite{thermo-Hehre-Book2003} it tends to be easier to obtain accurate reaction enthalpies of isodesmic reactions than for an isogyric reaction. However, only one isodesmic chemical reaction, \cref{rxn:H3PO4}, was identified leading to the formation of TBP and its derivatives. To verify the accuracy of the derived enthalpies of formation, also considered three isogyric \cref{rxn:HF,rxn:HCl,rxn:H2}, that include chemical compounds with known enthalpies of formation. 

\subsection{Choice of the atomic basis sets and the treatment of electron correlation}

\begin{figure*}[ht!]
\subfloat[\ce{H2MBP}]{\includegraphics[width=0.26\linewidth]{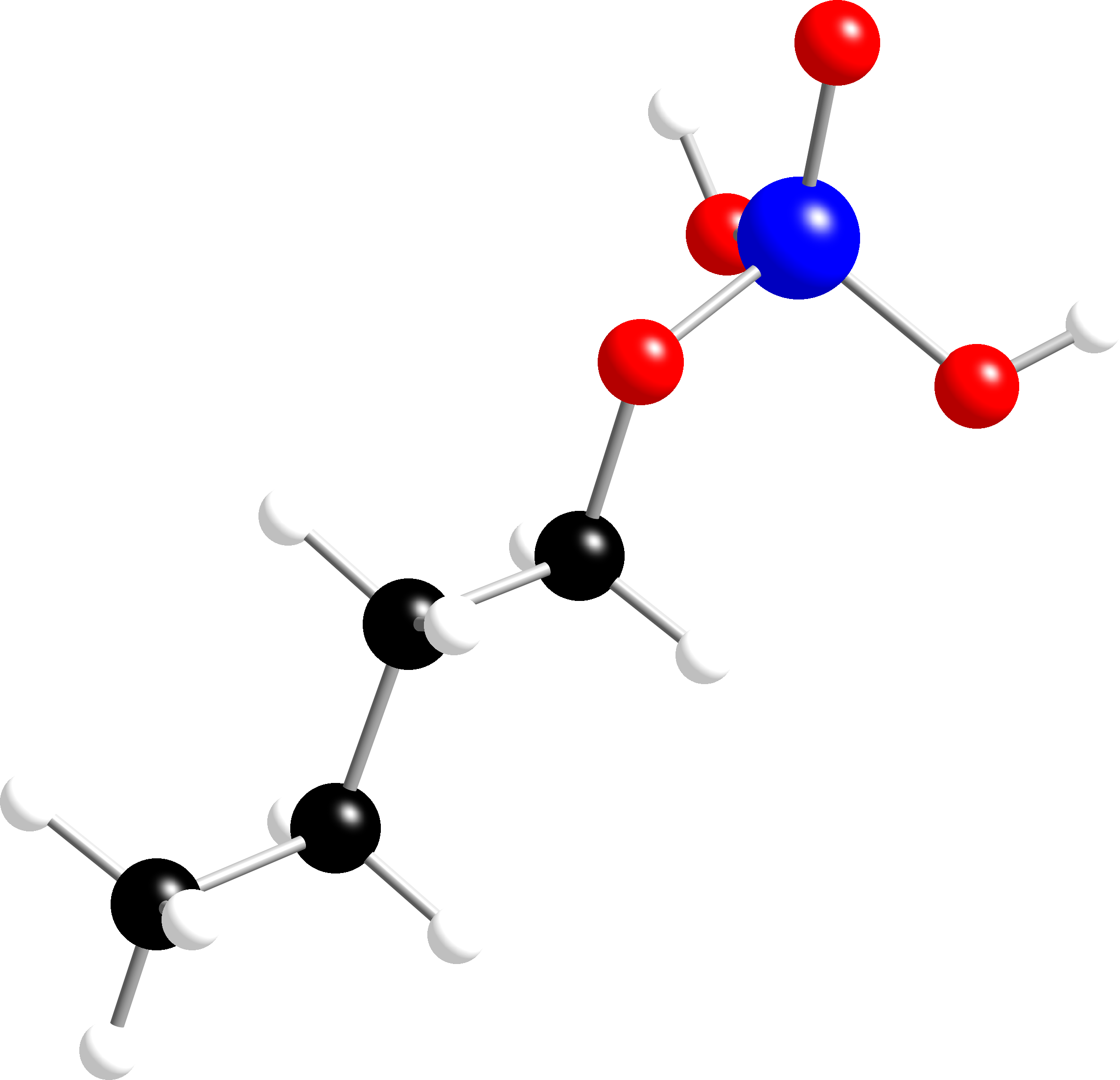}}
\subfloat[\ce{HDBP}]{\includegraphics[width=0.30\linewidth]{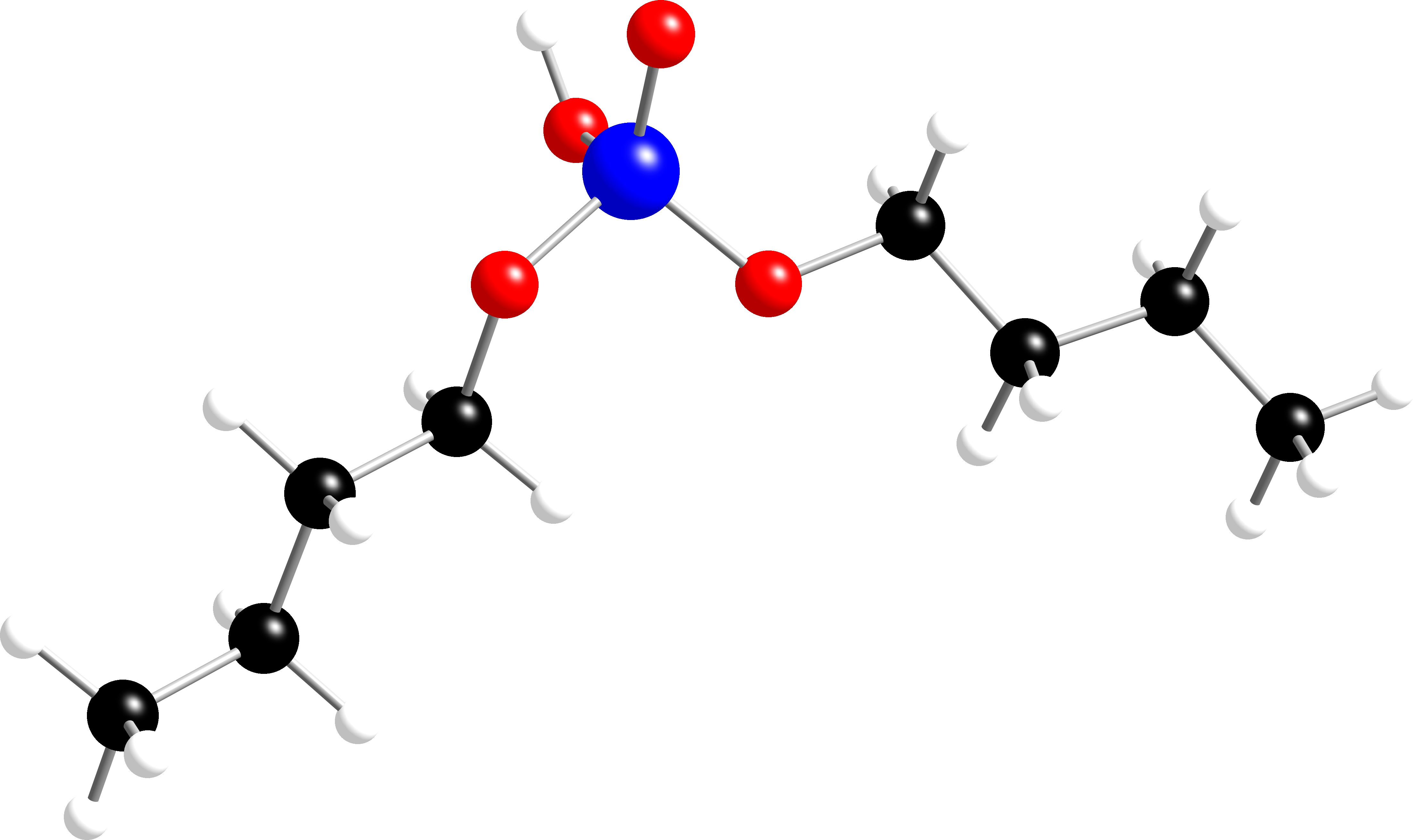}}
\subfloat[\ce{TBP}]{\includegraphics[width=0.30\linewidth]{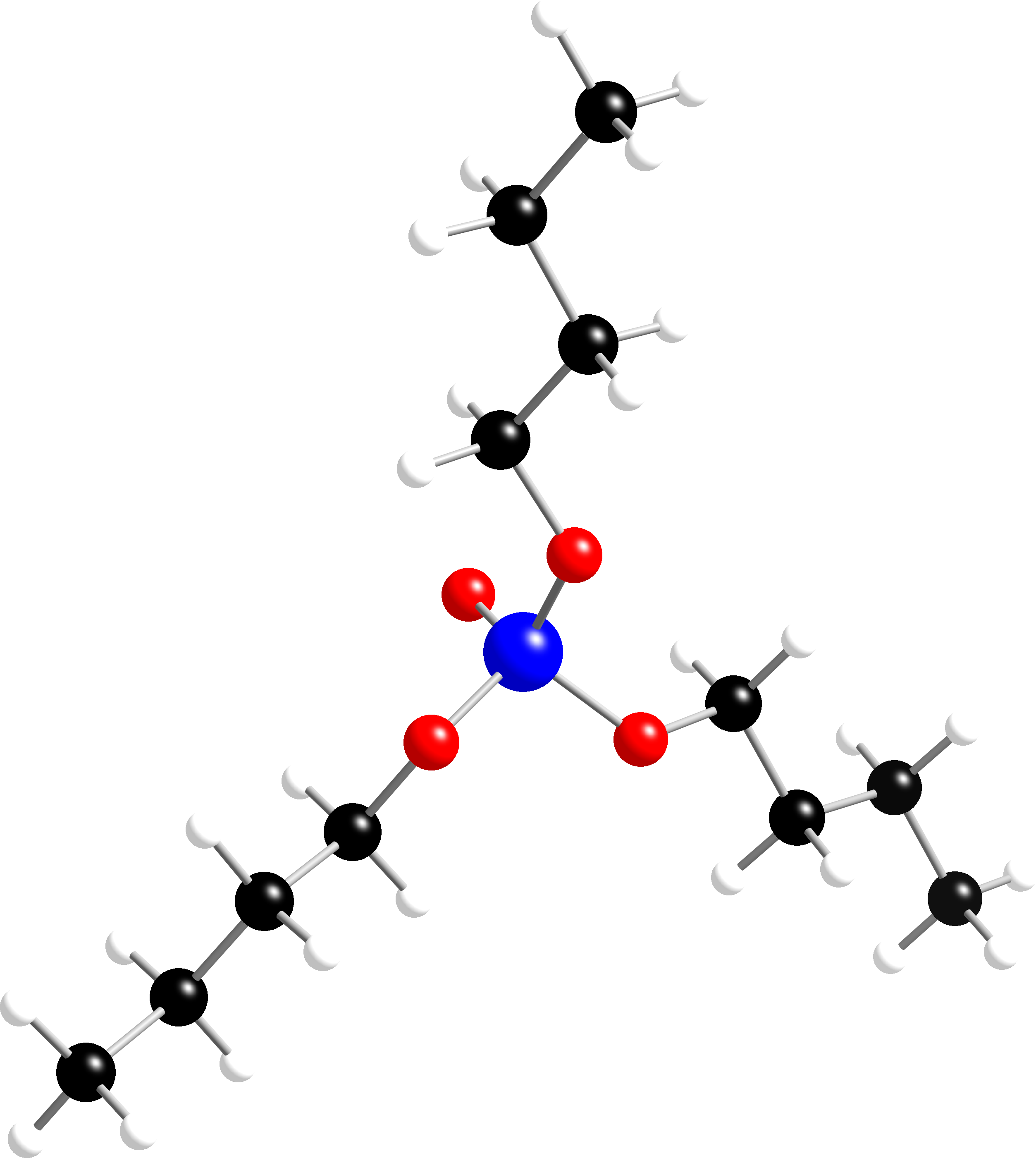}}
\caption{\label{fig:HnBP}Perspective views of the geometries of \ce{H2MBP}, \ce{HDBP} and \ce{TBP} optimized at the B3LYP-def2-TZVP level of theory. Hydrogen atoms are in white, carbon atoms in black, oxygen atoms in red, and phosphorous in blue.}

\end{figure*}

The optimal structures of all species were optimized in the gas phase without symmetry constraints using the B3LYP density functional.\cite{dft-Becke-JCP1993-98-5648,dft-Lee-PR1988-37-785,dft-Vosko-CJC1980-58-1200,dft-Stephens-JPC1994-98-11623} Triple-$\zeta$ valence basis sets def2-TZVP basis sets\cite{basis-Weigend-CPL1998-294-143,basis-Weigend-PCCP2005-7-3297} were used for all atoms. To describe butyl-phosphate \ce{H_{3-n}nBP(l)} molecules in the liquid phase, solvent effects have to be considered. However, we noted that the TBP solvent, treated by the COSMO continuum model,\cite{solvent-Klamt-JCSPT1993-799} induce only minor changes on the optimal structures of the molecules. Hence, gas-phase optimized geometries will be used throughout to compute the harmonic frequencies that enter, without being scaled, the vibrational partition functions at {\Tref} and 0.1~MPa. The latter are necessary to calculate the enthalpic and entropic contributions to the gas-phase energies. Perspective views of the optimal geometries of \ce{H2MBP}, \ce{HDBP} and \ce{TBP} are displayed on \cref{fig:HnBP}.

The derivation of unknown formation enthalpies requires to compute reaction enthalpies, which are composed of reaction energies and enthalpic corrections. The accuracy is mostly determined by the errors in the computed electronic reaction energies with respect to both the basis set (basis-set error) and the treatment of electron correlation (intrinsic error with respect to the exact energy), by considering the \cref{rxn:H3PO4,rxn:HCl,rxn:H2} leading to the formation of the smallest butyl phosphate \ce{H2MBP}. The other di and tributyl phosphate are difficult in this respect due to their size, which makes it impossible to consider huge basis sets very close to complete basis set (CBS) limit, at least for the correlated WFT methods, although large basis sets are much less prone to basis set superposition errors (BSSE). On the other hand, smaller basis sets have to be checked for the remaining basis set incompleteness error (BSIE) and BSSE. While the BSIE can be estimated with extrapolation schemes, the intramolecular BSSE is not directly accessible for the investigated formation reactions listed above.

We have used either the triple and quadruple-$\zeta$ valence basis sets of def2-type (def2-TZVP and def2-QZVP)\cite{basis-Weigend-CPL1998-294-143,basis-Weigend-PCCP2005-7-3297} to extrapolate to the CBS limit,\cite{basis-Neese-JCTC2011-7-33} or the sequence of augmented correlation consistent basis sets, aug-cc-pV$n$Z, from triple ($n$=3) to quintuple-$\zeta$ ($n$=5) quality with subsequent CBS extrapolation. Correlation effects are treated either with the B3LYP functional or with correlated wave-function theory (WFT) methods of increasing accuracy, namely MP2 and the single and double coupled cluster theory with inclusion of a perturbative estimation for triple excitation [CCSD(T)], the latter representing the ``gold standard''. All WFT calculations are performed with either the parallel resolution of the identity approximation,\cite{mrpt2-Hattig-JCP2000-113-5154,mrpt2-Hattig-PCCP2006-8-1159}  or density fitting correlated methods,\cite{prog-Werner-JCP2011-135-144116} with the appropriate atomic auxiliary basis functions,\cite{basis-Weigend-CPL1998-294-143,basis-Weigend-PCCP2005-7-3297} and the frozen-core approximation (chemical core, that is, only the valence electrons were correlated). For the aug-cc-pVnZ basis sets, extrapolation to the CBS limit was carried out according to the three-point exponential formula for HF energies $E_\text{HF}$:\cite{cbs-Feller-JCP1992-96-6104,cbs-Feller-JCP1993-98-7059}
\begin{equation}
\label{eq:CBS_HF}
E_\text{HF}(n) = E^{\text{CBS}}_\text{HF} + A \exp{(-Bn)},
\end{equation}
and a two-point extrapolation for the total DFT energies or the WFT correlation energies $E_\text{corr}$:\cite{cbs-Helgaker-JCP1997-106-9639}
\begin{equation}
\label{eq:CBS_corr}
E_\text{corr}(n) = E^{\text{CBS}}_\text{corr} + A n^{-3}.
\end{equation}

If calculations on the monobutyl-phosphate molecule (20 atoms) can be performed with "standard" (using canonical molecular orbitals) post-Hartree-Fock MP2 and CCSD(T) methods, their prohibitive scaling, $O(N^5)$ and $O(N^7)$ with $N$ the measure of the molecular size or the number of correlated electrons, renders calculations on the dibutyl and tributyl phosphate ligands unfeasible. However, the efficient local CCSD(T) (LCCSD(T)) method with density fitting approximation of the integrals offers a favorable $N$ scaling.\cite{prog-Werner-JCP2011-135-144116} To reach the highest accuracy in local coupled cluster treatments, Schwilk~{\etal}\cite{prog-Schwilk-JCP2015-142-121102} advocate to include all close pairs amplitudes in the LCCSD (\texttt{keepcls=2} option in MOLPRO.\cite{prog-molpro,prog-molpro-WIRES2012}) As discussed by Werner and Schütz,\cite{prog-Werner-JCP2011-135-144116} for reactions involving large molecules, one might observe significant differences between the results of local and canonical calculations, originating either from the local approximations or the larger BSSE effects in the canonical calculations. Therefore, it is recommended to add a correction $\Delta E_\mathrm{MP2(CBS)} = E_\mathrm{MP2(CBS)}-E_\mathrm{LMP2}$, to the LCCSD(T) energies, where $E_\mathrm{MP2(CBS)}$ is the CBS extrapolated canonical MP2 energy using aug-cc-pV$n$Z basis sets ($n$ = 3--5), and extrapolation formulas~\cref{eq:CBS_HF} and \cref{eq:CBS_corr}. The LMP2 and LCCSD(T) calculations are computed with triple-$\zeta$ basis sets. We will note this energy $\Delta E_\mathrm{LCCSD(T)(CBS^\ast)}$:

\begin{widetext}
\begin{equation}
\label{eq:CBS}
\Delta E_\mathrm{LCCSD(T)(CBS^\ast)} = E_\mathrm{LCCSD(T)} + E_\mathrm{MP2(CBS)}-E_\mathrm{LMP2}
\end{equation}
\end{widetext}

All calculations were performed with Turbomole\cite{prog-turbomole712} and MOLPRO.\cite{prog-molpro,prog-molpro-WIRES2012}

\subsection{Benchmarking basis sets and correlation methods for the \ce{H2MBP} formation \cref{rxn:H3PO4,rxn:HCl,rxn:H2}}
\label{sec:methods}

\begin{table}[h!]
\caption{\label{tabmethod-comp}
Reaction energies $\Delta_rE$ in \SI{}{\kJ\per\mol} for \ce{H2MBP} formation \cref{rxn:H3PO4,rxn:HCl,rxn:H2}, using different methods and basis sets.}

\begin{tabular}{ll*3{l}}
\hline
{Method}&{Basis set} &\multicolumn{1}{c}{$\Delta_rE$ (\ref{rxn:H3PO4})} &\multicolumn{1}{c}{$\Delta_rE$  (\ref{rxn:HCl})} &\multicolumn{1}{c}{$\Delta_rE$(\ref{rxn:H2})} \\
\hline
B3LYP&{def2-TZVP}& 0.0& -131.1 & -105.5\\
B3LYP&{def2-QZVP}&-2.3&-124.2& -103.3\\
B3LYP &{CBS}&-4.0&-119.3& -101.8\\
\hline
B3LYP&{def2-QZVPP}&-2.3&-124.2& -103.3\\
\hline
B3LYP&{aug-cc-pVTZ}&-3.9&-108.9& -51.7\\
B3LYP&{aug-cc-pVQZ}&-3.7&-116.3& -78.1\\
B3LYP &{aug-cc-pV5Z}&-3.2&-121.1& -97.1\\
B3LYP&{CBS}&-2.8&-126.0& -117.0\\

\hline
MP2&{def2-TZVP}&-8.0&-135.7 & -202.5\\
MP2&{def2-QZVP}&-10.0&-119.0& -155.9\\
MP2 &{CBS} &-11.4&-106.7 & -121.9\\
\hline
MP2&{aug-cc-pVTZ}&-13.8&-112.0& -110.5\\
MP2&{aug-cc-pVQZ}&-12.3&-112.8& -141.5\\
MP2&{aug-cc-pV5Z}&-11.6&-113.3& -161.6\\
MP2&{CBS}&-11.1&-111.1& -177.9\\

\hline
LCCSD(T)&CBS$^{\ast}$&-10.3&-124.8& -136.0\\
\hline
\end{tabular}
\end{table}

Reaction~\ref{rxn:H3PO4} is isodesmic, implying that the type of chemical bonds broken in the reactants are the same as the type of bonds formed in the reaction products. Such type of reactions are used preferentially in thermochemistry to calculate accurate enthalpies of formation. As we mentioned previously, as only one isodesmic reaction was found, we considered as well three non-isodesmic reactions that are isogyric: both \cref{rxn:HCl,rxn:HF} retain the \ce{P=O} bond, while \cref{rxn:H2} does not. We can thus expect better error cancelations in the former two than in the latter. \cref{tabmethod-comp} gathers the computed energies for \cref{rxn:H3PO4,rxn:HCl,rxn:H2}, leading to the formation of \ce{H2MBP} with different basis sets and DFT and WFT methods.\\

Let us begin by scrutinizing the results for \cref{rxn:H3PO4}. Both def2-basis and aug-cc-pV$n$Z basis set families yield very similar CBS values with the B3LYP functional. This is also true at the MP2 level, though the CBS reaction energies are about \SI{-7.5}{\kJ\per\mol} more exothermic that at the B3LYP level. The LCCSD(T) agrees within  \SI{0.8}{\kJ\per\mol} with the MP2 one. This suggests that electron correlation effects essentially cancel out in this isodesmic reaction, and confirms that the fact that error compensations in such reactions are optimal.

With respect to the isogyric reactions, we first focus on \cref{rxn:HCl}. With the B3LYP functional, the reaction energy exhibits an unexpected strong basis set dependence with the def2 basis sets. The reaction energies are not converged with the largest basis set QZVP. Extending the def2-QZVP basis set with extra polarization functions (def2-QZVPP) has no effect on the computed values. The other surprising observation is that the aug-cc-pV$n$Z basis sets also converge slowly to the basis set limit, together with the fact that both def2- and cc-type basis sets lead to CBS values that differ by \SI{6.7}{\kJ\per\mol}. Whenever large basis set dependences are observed at the DFT level, one may suspect large BSSE even with very extended atomic orbital basis that results from numerical instabilities of the DFT functionals, as discussed recently by Hansen~{\etal}\cite{basis-Hansen-C2014-3-177} This makes us conclude that BSIE and BSSE effects are crucial with the B3LYP functional for the various reactants and products of \cref{rxn:HCl,rxn:H2}. The electron correlation effects appear to be significantly different in the two sides of the two reaction equations, thus making error compensations not favorable, and forcing us to push the treatment of dynamic correlation up to the CBS limit.

At the MP2 level, while the def2 basis sets reaction energy values converge very slowly to its CBS limit, the aug-cc-pV$n$Z basis sets are far less prone to BSIE for \cref{rxn:HCl}, being only \SI{0.9}{\kJ\per\mol} apart from the CBS level with triple-$\zeta$ basis set. MP2/aug-cc-pV$n$Z energies for \cref{rxn:H2} converge more slowly to its CBS limit. Because the def2-CBS limit significantly differs from the aug-cc-pV$n$Z one, we cross-checked these results with relativistic (Douglas Kroll Hess, DK) correlation consistent basis sets (aug-cc-pV$n$Z-DK) and confirmed the non-relativistic aug-cc-pV$n$Z values. We thus infer that the very different bonding around the phosphorous atom in the reactant \ce{PH3} (three single \ce{P-H} bonds) and in the product \ce{H2MBP} (one double \ce{P=O} bond, three single \ce{P-O} bonds) cannot be accurately captured by def2-basis sets.

To push the treatment of electron correlation further, we performed LCCSD(T) calculations extrapolated to the CBS limit as explained earlier in this section. The resulting reaction energies estimated to the CBS limit are \SI{-124.8}{\kJ\per\mol} and \SI{-136.0}{\kJ\per\mol} for \cref{rxn:HCl} and \cref{rxn:H2}, respectively. Though we cannot compare this value to experimental values, we trust LCCSD(T)/CBS results as the best estimate for CCSD(T)/CBS accuracy, and we will see in the next section that this level of theory helps us deriving accurate standard enthalpies of formation.

\subsection{COSMO-RS method to estimate solvent contributions to thermodynamic quantities}
\label{sec:COSMO-RS}
Whenever the molecules considered are in the liquid phase, we have to estimate their thermodynamic properties in the corresponding phase. While most continuum solvent models only estimate Gibbs Free energy of solvation, the Conductor-like screening model-Real Solvents (COSMO-RS) model\cite{solvent-Eckert-ACI2002-48-369} is the only one that allows us to access enthalpies of solvation, which in a pure solvent are strictly equal to the opposite of the enthalpies of vaporization enthalpy, or for a solvated molecule can be extracted as the derivative of the solvation free energy by finite differences with two different temperatures (\SI{25}{\celsius}, \SI{30}{\celsius}).\cite{solvent-Diedenhofen-PCCP2007-9-4653} In order to obtain reliable thermophysical property data with COSMO-RS predictions, accurately optimized structures with corresponding gas-phase and COSMO state (ideal dielectric continuum by using COSMO with $\epsilon_r=\infty$) energies are necessary.\cite{solvent-Diedenhofen-PCCP2007-9-4653} These are performed with the BP functional of the density\cite{dft-Becke-PR1988-38-3098,dft-Perdew-PR1986-33-8822,dft-Perdew-PR1986-34-7406} and def-TZVP basis sets.\cite{basis-Schafer-JCP1994-100-5829}  All statistical COSMO-RS thermodynamic calculations\cite{solvent-Eckert-ACI2002-48-369} were performed with COSMOtherm,\cite{prog-cosmotherm} using the parameter file BP TZVP``C30\_1701''. 

\begin{table}
\caption{\label{tab:Hvap}
COSMO-RS vaporization enthalpies of all relevant species in \SI{}{\kJ\per\mol}, together with experimental values when available.}
\begin{tabular}{ll*2{l}l}
\hline
{Molecule} & {liquid phase} & \multicolumn{2}{c}{{\Hvap}({\Tref})}\\
&&{COSMO-RS} &{Exp.}& {Ref.}\\
\hline
{\ce{C4H9OH}} & {\ce{C4H9OH}}&58.8 & \num{52.34(2)} &\citenum{physchem-Chickos-TA1995-249-41}\\
{\ce{H2O}} & {\ce{H2O}}& 49.2 &\num{43.99(7)}&\citenum{physchem-Marsh-Book1987}\\
{\ce{TBP}}& {\ce{TBP}}&85.6& 64.4 &\citenum{thermo-Burger-Book1990}\\
{\ce{HDBP}}& {\ce{TBP}}& 118.5&&\\
& {\ce{HDBP}} & 119.3 &\\
{\ce{H2MBP}}&{\ce{TBP}}&159.5 &&\\
& {\ce{H2MBP}} & 147.1 &\\
{\ce{H3PO4}}&{\ce{H3PO4}}&144.7 &&\\


\hline
\end{tabular}
\end{table}%

To evaluate the accuracy of COSMO-RS out-coming thermodynamic quantities, we first start by comparing the calculated enthalpies of vaporization {{\Hvap}} to available experimental values for butanol, water and TBP. The results reported in~\cref{tab:Hvap} shows that the COSMO-RS {\Hvap} values overestimate by 6.5 and \SI{5.2}{\kJ\per\mol} the experimental ones for butanol and water, respectively. For TBP, it overshoots by \SI{21}{\kJ\per\mol}. Such deviations are within the range of the reported range of accuracy.\cite{solvent-Diedenhofen-PCCP2007-9-4653,solvent-Schroder-FPE2014-370-24}  For \ce{H3PO4}, there does not exist an experiment value for the enthalpy of formation in the gaseous phase. Thus, we have calculated the  enthalpy of vaporization {{\Hvap}} for \ce{H3PO4} using COSMO-RS and deduced the enthalpy of formation in the gaseous phase from that in the liquid phase and the calculated enthalpy of vaporization. This leads us to a gas-phase {\Hform}({\Tref}) value of \SI{-1126.96\pm2.9}{\kJ\per\mol}, in very good agreement with the value of \SI{-1141.4}{\kJ\per\mol} calculated at the CCSD(T)-CBS level by Alexeev~{\etal}\cite{actinide-Alexeev-IJQC2005-102-775}
 

We therefore conclude that COSMO-RS can predict solvation contributions to a reaction energy with good confidence and help us to determine the unknown formation enthalpies of \ce{H2MBP}, \ce{HDMP} and \ce{TBP} in the TBP organic phase.

\section{Results and Discussion}
\label{sec:results-discussion}
\begin{table}
\begin{center}
\caption{\label{tab:Hf}{\Hform}({\Tref}) of TBP, HDBP, and \ce{H2MBP} in the gaseous phase calculated from the isodesmic \cref{rxn:H3PO4} and the average of the three isogyric \cref{rxn:HF,rxn:HCl,rxn:H2} listed in~\cref{tab:Hf:isogyric}. $\overline{\Delta_fH^{\ominus}}$({\Tref}) represent the average between these two data sets. The standard deviations {$\mathrm{\Delta}${\Hform}} includes the uncertainty on the average value as well as the experimental error bars. Values in the  liquid state are given by {\Hform}({\Tref})= $\overline{\Delta_fH^{\ominus}}$({\Tref})-{\Hvap}({\Tref}). All results are in \si{\kJ\per\mol}.}
\begin{tabular}{cc*2{l}}
\hline
Species&{Reaction} &{{\Hform}({\Tref})}&{$\mathrm{\Delta}${\Hform}}\\
\hline
\multirow{3}{*}{TBP$_{(g)}$}&{\cref{rxn:H3PO4}}&-1278.4&18.0\\
&{\cref{rxn:HF}}--{\cref{rxn:H2}}&-1285.0 &23.8\\
&{$\overline{\Delta_fH^{\ominus}}$({\Tref})  =}& -1281.7&\multirow{2}{*}{24.4}\\
\cline{1-3}
\multirow{1}{*}{TBP$_{(l)}$}&&-1367.3\\
\hline
\multirow{3}{*}{HDBP$_{(g)}$}&{\cref{rxn:H3PO4}}&-1226.1&13.0\\
&{\cref{rxn:HF}}--{\cref{rxn:H2}}&-1232.7&19.0\\
&{$\overline{\Delta_fH^{\ominus}}$({\Tref}) =}& -1229.4&\multirow{2}{*}{19.6}\\
\cline{1-3}
\multirow{1}{*}{HDBP$_{(l)}$}&&-1348.7\\
\hline
\multirow{3}{*}{H$_2$MBP$_{(g)}$} &{\cref{rxn:H3PO4}}&-1173.4&7.9\\
&{\cref{rxn:HF}}--{\cref{rxn:H2}}&-1180.0&14.3\\
&{$\overline{\Delta_fH^{\ominus}}$({\Tref}) = }& -1176.7&\multirow{2}{*}{14.8}\\
\cline{1-3}
\multirow{1}{*}{H$_2$MBP$_{(l)}$}&&-1323.8\\
\hline
\end{tabular}

\end{center}
\end{table}%

First, the standard enthalpies of formation of  \ce{H2MBP}, HDBP and TBP species are determined in gaseous state from  \cref{rxn:H3PO4,rxn:HF,rxn:HCl,rxn:H2}.  Electronic energies of  involved species are computed with the LCCSD(T)/CBS approach whereas the vibrational contributions are calculated at the B3LYP/def2-TZVP level (like geometry optimization) to ultimately derive the enthalpies of reaction. Note that we are confident in the quality of the partition functions and computed vibrational spectra, because of the estimated values of enthalpy changes, entropies and heat capacities reported in Tables~S3 and S5 of the supplementary material agree excellently with experimental data. To transform  the standard enthalpy of reaction towards the standard enthalpy of formation, the standard heats of formation at {\Tref} of key species (See~\cref{tab:Hfdata}) are taken into account. 
Secondly the computation of solvation enthalpies of target species is performed with the COSMO-RS method, as introduced in \cref{sec:COSMO-RS}. Hence the combination of two last computed values leads the standard enthalpy of formation in the TBP solvent.

\subsection{Heat of formation of TBP}

To validate our approach in computing enthalpies of formation, the type of the chosen chemical reaction might be critical. Inspection of~\cref{tab:Hf} reveals that the three isogyric \cref{rxn:HF,rxn:HCl,rxn:H2} yields values of enthalpies of formation that are very close to each other. For TBP in the gaseous state, the average over these three reactions (See~Table~S2 in the supplementary material) is \SI{-1285.0\pm23.8}{\kJ\per\mol}, which compares extremely well with the value derived with the isodesmic \cref{rxn:H3PO4}, \SI{-1278.4\pm18.0}{\kJ\per\mol}. The same is true for the formation enthalpies of \ce{HDBP} and \ce{H2MBP} (Comparison of the isogyric averages in~Table~S2 of the supplementary material and the isodesmic values in~\cref{tab:Hf}). We thus decided to define our best estimates from the average of \cref{rxn:H3PO4} and the isogyric \cref{rxn:HF,rxn:HCl,rxn:H2}, $\overline{\Delta_fH^{\ominus}}$({\Tref}), with a standard deviation {$\mathrm{\Delta}${\Hform}} that includes the uncertainty on the average value as well as the experimental error bars listed in~\cref{tab:Hfdata}. 

For gaseous TBP, we predict a value of \SI{-1281.7\pm24.4}{\kJ\per\mol}, which is in line with the value obtained from the group parameter correlation method, \SI{-1310}{\kJ\per\mol}.\cite{Kertes1972796} As there is no other data available for {\Hform} of TBP, our data could be considered as the most accurate one.

The {\Hform} of TBP in the liquid state are obtained from data in the gaseous state and its corresponding heat of solvation  (\SI{-85.6}{\kJ\per\mol}, see~\cref{tab:Hvap} in~\cref{sec:COSMO-RS}), the value \SI{-1367.3\pm24.4} {\kJ\per\mol} is obtained. The experimental data are scattered, thereby the heats of combustion of TBP obtained by  bomb calorimeter were reported in only three experimental studies. \cite{actinide-Starostin-IANSSK1966-15-1255,actinide-Kindle-techreport1974,physchem-Erastov-RJPC1992-66-2591} The derived standard enthalpies of formation exhibit a large uncertainty over a range of about \SI{200}{\kJ\per\mol}. These disagreements shall be assessed from the details given by the authors because the calorimetry requires a very accurate analysis regarding correction factors  of each experimental step, therefore, a study lacking all details has to be cautiously considered. 

Kindle \cite{actinide-Kindle-techreport1974} investigated the heat release of the TBP combustion for a safety analysis of flammability. Initially, these data were not intended for the evaluation of  thermodynamic quantities. Nevertheless, an estimation is possible by considering a complete combustion, as performed in a review of TBP properties:\cite{thermo-Burger-Book1990} \SI{-1250\pm69}{\kJ\per\mol}. But this experiment is tainted by unburnt TBP as well as soot formation. Then author does not mention any analysis of the purity of the TBP sample and the post-combustion products are not fully investigated. From these statements, the value extracted from Kindle's report is rejected even if the calorimeter calibration, as well as explanation of each correction factor, have been perfectly performed. 

Two other investigations \cite{actinide-Starostin-IANSSK1966-15-1255,physchem-Erastov-RJPC1992-66-2591} used a  bomb  calorimeter with an isothermal shell. However, as opposed to Starostin,\cite{actinide-Starostin-IANSSK1966-15-1255} Erastov and Tarasov\cite{physchem-Erastov-RJPC1992-66-2591} has employed a rotating bomb. The main benefit of this facility is  to ensure a complete combustion process and a good thermal transfer between sample zone and shell. To go further, the analyses carried out by Starostin \cite{actinide-Starostin-IANSSK1966-15-1255} do not show the hydrolysis of pyrophosphoric acid certainly due to the use of static shell. In addition, as the both experiments are similar, the two reported net heat released can be compared; Starostin has measured the lowest heat release highlighting also the consequence of static bomb. These findings suggest that Erastov and Tarasov's measurements \cite{physchem-Erastov-RJPC1992-66-2591} are the most relevant. Otherwise, Erastov and Tarasov have purified the TBP followed by a check of impurities content (moisture $<$~0.1\%mol and no impurity). Then each sample is sealed in a container made of polyester to prevent any reaction with the environment before combustion run. Furthermore, a complete analysis at the end of runs have been performed in order to fully examine the composition of by-products. To conclude, the standard heat of formation extracted from Erastov and Tarasov's investigations is considered as the most accurate and equal to \SI{-1382\pm12}{\kJ\per\mol}.

Finally, our theoretical standard enthalpy of formation is in agreement with the selected experimental data. The discrepancy between them lies below \SI{14.7}{\kJ\per\mol}, therefore, validating our quantum chemistry approach and its ability in predicting thermochemical data. Thus, it gives confidence for applications to  degraded products, \ce{HDBP(l)} and \ce{H2MBP(l)}.   

\subsection{Heat of formation of HDBP and \ce{H2MBP} and investigation of the hydrolysis reactions}

\begin{table*}[htb!]
\begin{center}
\caption{\label{tab:hydrolysis}
Ab initio $\mathrm{LCCSD(T)(CBS^\ast)}$ Gibbs reaction energies for the hydrolysis reactions \ce{H_{3-n}nBP(l) + H2O(l) -> H_{4-n}(n-1){BP}(l) + C4H9OH(l)} in the TBP and water solvents at {\Tref} and \SI{363.15}{\kelvin} (\SI{90}{\celsius}). All values are in \si{\kJ\per\mol}.}

\begin{tabular}{l*3{l}}
\hline
Reaction & {{\Hreac}({\Tref})} & {{\Greac}({\Tref})} & {{\Greac}(\SI{363.15}{\kelvin})} \\
\hline
& \multicolumn{3}{c}{TBP liquid} \\
\cline{2-4}
\ce{TBP(l) + H2O(sln) -> HDBP(sln) + C4H9OH(sln)} 
& -27.4& 0.4 & 5.7 \\

\ce{HDBP(sln) + H2O(sln) -> H2MBP(sln) + C4H9OH(sln)}
& -32.6& -2.6 & 3.1 \\

\ce{H2MBP(sln) + H2O(sln) -> H3PO4(sln) + C4H9OH(sln)}
& -35.8 & -6.8 & -1.2 \\
\cline{1-4}
\ce{TBP(l) + 3 H2O(sln) -> H3PO4(sln) + 3 C4H9OH(sln)} 
&-95.8 & -9.0 & 7.7 \\ 
\hline
& \multicolumn{3}{c}{water} \\
\cline{2-4}
\ce{TBP(aq) + H2O(l) -> HDBP(aq) + C4H9OH(aq)} 
&  -21.6 & 0.6 & 4.9\\

\ce{HDBP(aq) + H2O(l) -> H2MBP(aq) + C4H9OH(aq)}
& -23.7 &-1.8 & 2.5\\

\ce{H2MBP(aq) + H2O(l) -> H3PO4(aq) + C4H9OH(aq)}
& -22.7 &-3.8 & -0.1\\
\cline{1-4}
\ce{TBP(aq) + 3 H2O(l) -> H3PO4(aq) + 3 C4H9OH(aq)} 
&-68.0 & -5.0 & 7.2 \\ 
\hline
\end{tabular}
\end{center}
\end{table*}

The calculated values of standard enthalpy of formation for both degraded products are reported in~\cref{tab:Hf}. In the liquid state, {\Hform} of HDBP and \ce{H2MBP} are \SI{-1348.7\pm19.6}{} and \SI{-1323.8\pm14.8}{\kJ\per\mol}, respectively. Again, there is no available data to compare these results. 

To go further from these standard heats of formation, the hydrolysis phenomena can be analyzed in both TBP and water solvents \cref{tab:hydrolysis} summarizes the heat and Gibbs energies of reaction for different temperatures.  The hydrolysis process (see reactions in~\cref{tab:hydrolysis}) corresponds to the cleavage of \ce{P-O} and \ce{O-H} bonds together with the formation of the same bonds regarding each of the three intermediate steps. Consequently, the step-wise heats of reaction and the overall one are expected to be nearly thermoneutral. Obtained heats of reaction are weakly exothermic in both solvents which is consistent with the later assumption. The  trend is that the larger butyl phosphate, the lower exothermic its degradation. Nevertheless, it is commonly admitted that TBP slowly degrades by an acid catalyzed hydrolysis even at room temperature \cite{Burger58} leading to a negative Gibbs energy of reaction. However, our data sometimes highlight slightly positive value of Gibbs energy of reaction meaning an unfavorable process in contradiction with the TBP knowledge.  At room temperature, whatever the solvent, there is only the first hydrolysis step that is weakly endergonic but the value is smaller than the modeling accuracy of COSMO-RS. The overall hydrolysis reaction remains a favorable process. At temperature just below boiling point, the Gibbs reaction energies become more endergonic except for hydrolysis of mono-butyl phosphate. 
These disagreements can be explained by the fact that our theoretical model reaches its accuracy limits to predict a weak thermal effect. In addition,  the entropy of such species is very sensitive  data and requires a detailed vibrational analysis of molecular motions in solution. Therefore its contribution to Gibbs reaction energy may be responsible too. Thereby, the TBP, HDBP and \ce{H2MBP} hydrolysis should moderately contribute to the heat release during a thermal runaway.

\section{Conclusions}
In this paper, we have established a methodology to predict the thermodynamic parameters of \ce{TBP} and its degradation products \ce{HDBP} and \ce{H2MBP} in both gas and liquid phases. Basis set and solvent effects are considered and incorporated into a composite thermochemistry scheme based on B3LYP/def2-TZVP geometries and partition functions, and LCCSD(T) extrapolated to the complete basis set limit for calculating accurate reaction enthalpies that are used to determine heats of formations for the butyl-phosphate molecules. The excellent similarities of the derived standard enthalpies of formation irrespective to the chosen formation reaction, makes us propose that the uncertainty in the herewith reported {\Hform} is in the range of \SI{20}{\kJ\per\mol} in the gas phase. In the solvent, the hereafter quoted standard deviations have to be complemented by the error induced by the COSMO-RS solvent model (about ~\SI{22}{\kJ\per\mol} for \ce{TBP}). This work makes us propose the value \SI{-1367.3\pm24.4}{\kJ\per\mol} for the heat of formation ({\Tref}) of \ce{TBP}, confirming the value of \SI{-1382\pm12}{\kJ\per\mol} proposed by Erastov and Tarasov\cite{physchem-Erastov-RJPC1992-66-2591} In the absence of experimental data for \ce{HDBP} and \ce{H2MBP}, this work predicts their heats of formation in the liquid phase to be  \num{-1348.7\pm19.6} and  \SI{-1323.8\pm14.8}{\kJ\per\mol}, respectively. These data allow us to predict that the complete hydrolysis of \ce{TBP} is clearly exothermic.

\section{Supplementary material}
See supplementary material for 1) details on the estimate of the uncertainty for the {\Hform}({\Tref}) of \ce{H3PO4(l)}; 2) Energies and enthalpic contributions for ~\cref{rxn:H3PO4,rxn:HF,rxn:HCl,rxn:H2}; 3) Enthalpies of formation derived from the isogyric reactions; 4) Molecular thermodynamics data for the studied systems.

\section{Acknowledgments}
The members of the PhLAM laboratory acknowledge support from the CaPPA project (Chemical and Physical Properties of the Atmosphere) that is funded by the French National Research Agency (ANR) through the PIA (Programme d'Investissement d'Avenir) under contract``ANR-11-LABX-0005-0'' and by the Regional Council ``Hauts de France'' and the  "European Funds for Regional Economic Development" (FEDER) through the Contrat de Projets Etat-Région (CPER) CLIMIBIO (Changement climatique, dynamique de l’atmosphère, impacts sur la biodiversité et la santé humaine). We also acknowledge financial support from Slovak grants APVV-15-0105 and VEGA 1/0465/15, the computational resources of HPC Cluster of Slovak University of Technology (projects ITMS 26230120002 and 26210120002) and funding by the ERDF under the project "University Scientific Park Campus MTF STU - CAMBO" ITMS: 26220220179.

\bibliography{TBP} 

\end{document}